\def\tr{{\rm tr}\,}

\def\b{\bibitem}
\def\bml{\begin{mathletters}}
\def\eml{\end{mathletters}}
\def\be{\begin{equation}}
\def\ee{\end{equation}}
\def\bea{\begin{eqnarray}}
\def\eea{\end{eqnarray}}
\documentstyle[aps,prl,epsf,floats]{revtex}

\begin{document}
\def\SNG{{\em Physical Review Style and Notation Guide}}
\def\LUG {{\em \LaTeX{} User's Guide \& Reference Manual}}
\def\btt#1{{\tt$\backslash$\string#1}}%
\def\REVTeX{REV\TeX}
\def\AmS{{\protect\the\textfont2
        A\kern-.1667em\lower.5ex\hbox{M}\kern-.125emS}}
\def\AmSLaTeX{\AmS-\LaTeX}
\def\BibTeX{\rm B{\sc ib}\TeX}
\twocolumn[\hsize\textwidth\columnwidth\hsize\csname@twocolumnfalse%
\endcsname
\title{Fluctuation-Driven Quantum Phase Transitions in Clean Itinerant 
       Ferromagnets}
\author{D. Belitz}
\address{Department of Physics, and Materials Science
         Institute, University of Oregon, Eugene, OR 97403}
\author{T.R. Kirkpatrick}
\address{Institute for Physical Science and Technology, and Department 
         of Physics, University of Maryland, College Park, MD 20742.}
\date{\today}
\maketitle
\begin{abstract}
The quantum phase transition in clean itinerant ferromagnets is analyzed. 
It is shown that soft particle-hole modes invalidate Hertz's mean-field 
theory for $d\leq 3$. A renormalized mean-field theory predicts a 
fluctuation-induced first order transition for $1<d\leq 3$, whose stability 
is analyzed by renormalization group techniques. Depending on microscopic 
parameter values, the first order transition can be stable, or be pre-empted 
by a fluctuation-induced second order transition. The critical behavior at 
the latter is determined. The results are in agreement with recent experiments.
%
%
\end{abstract}
\draft
]

One of the most common and basic examples of a phase transition is the
paramagnet-to-ferromagnet transition in metals, e.g., iron and nickel.
These elements have high Curie temperatures, on the order of $1,000$ K.
Examples of itinerant ferromagnets with much lower Curie temperatures,
on the order of tens of K, include MnSi\cite{MnSi}, ZrZn$_2$\cite{MnSi,ZrZn},
and UGe$_2$\cite{UGe}. In the latter
materials, the Curie temperature can be suppressed to zero by the
application of hydrostatic pressure. This allows for an experimental
investigation of the ferromagnetic quantum phase transition, which takes
place at zero temperature as a function of a non-thermal control parameter,
in this case, pressure. Another example is Ni$_x$Pd$_{1-x}$\cite{NiPd}, 
where the control parameter is the nickel concentration. 

The ferromagnetic transition in metals was also the subject of the earliest
theoretical studies of quantum phase transitions. In an important paper,
Hertz\cite{Hertz} concluded that the critical behavior should be 
mean field-like in all dimensions $d>1$. This is because, in quantum
statistical mechanics, statics and dynamics are coupled. As a result,
the quantum phase transition in $d$ dimensions is related to its classical
counterpart in $d+z$ dimensions, with $z$ the dynamical critical exponent.
Since simple theories suggest $z=3$ for clean itinerant ferromagnets, and 
since the classical Heisenberg ferromagnet has an upper critical dimension 
$d_{\rm c}^+=4$, this argument suggests $d_{\rm c}^+ = 1$ for the quantum 
transition.

It is now known, mostly through studies of the corresponding problem in the
presence of quenched disorder, that the above argument is in general not
correct. The basic physical reason is the existence of soft modes, particle-hole
excitations in the case of itinerant ferromagnets, that couple to the order
parameter and preclude the construction of a Hertz-type Landau-Ginzburg-Wilson 
(LGW) theory entirely in terms of the order parameter. These
soft modes lead to time scales in addition to the critical one, and hence to
multiple exponents $z$. This in turn leads to an instability of the mean-field
fixed point that is not apparent in a power-counting analysis of the LGW
theory. In the disordered case, the net result is an upper critical dimension
$d_{\rm c}^+ = 4$ (instead of $d_{\rm c}^+ = 0$ in Hertz theory), with 
non-mean field (and non-power law) critical behavior in 
the physically most interesting dimension $d=3$\cite{us_fm_dirty}.
In the clean case, an analogous analysis of the stability of Hertz's fixed
point shows that the upper critical dimension is
$d_{\rm c}^+ = 3$. In $d=3$ a generalized Landau theory predicts a first order 
transition due to an $m^4\ln m$ term in the Landau free energy, with $m$ 
the magnetization\cite{us_letter}. For $1<d<3$ this theory also predicts
a first order transition. Sufficiently high temperature or 
disorder lead to an analytic Landau free energy and render the transition 
second order. 

This theoretical situation is at best in partial agreement with experiments. 
The existing theory predicts that the transition should always be of first
order in sufficiently clean materials at sufficiently low 
temperatures\cite{hypothesis_footnote}. If
temperature or disorder drives it second order, then the predicted critical
exponents are the classical Heisenberg exponents, or the strongly non-mean 
field exponents of Ref.\ \onlinecite{us_fm_dirty}, respectively. 
Experimentally, the transition in MnSi and UGe$_2$ at low 
temperatures is indeed observed to be of first order\cite{MnSi,UGe}, 
but in ZrZn$_2$ it is of second order even in very clean samples at 
very low temperatures\cite{ZrZn}, and the same is true in NiPd\cite{NiPd}. 
Furthermore, the critical behavior in NiPd was found to be
mean field-like to within the experimental accuracy,
in good agreement with the predictions of Ref.\ \onlinecite{Hertz}. This 
is suprising, given the above conclusion that Hertz theory cannot be
correct in $d=3$.

In this Letter we show that the nature of the clean ferromagnetic quantum 
phase transition is determined by physical effects that had not previously been 
recognized, and that taking these effects into account removes the
above discrepancies between theory and experiment. We will use
renormalization group (RG) techniques to show the following. (1) The first order
transition can be understood as a fluctuation-induced first order transition.
(2) For certain microscopic parameter values the first order transition is
unstable with respect to a fluctuation-induced second order transition. (3) The
upper critical dimension is $d_{\rm c}^+=3$. (4) In the second order case, the
critical behavior in $d=3$ is given by mean-field exponents with 
logarithmic corrections, and in $d<3$ it can be controlled by means of a 
$3-\epsilon$ expansion. We will present our results first, and then sketch 
their derivation and explain their physical origin.

If the bare value of the quartic coefficient in the Landau free energy is
sufficiently small (in a sense to be specified below), one finds the first
transition discussed in Ref.\ \onlinecite{us_letter}. However, if the bare
quartic coefficient is sufficiently large, the transition is of second order.
In $d=3$, the critical behavior is mean field-like with logarithmic corrections
to scaling. Specifically, the paramagnon propagator in the critical regime in
the paramagnetic phase has the form
\bml
\label{eqs:1}
\be
{\cal M}(k,\Omega_n) = 1/(t + a(k)\,k^2 + \vert\Omega_n\vert/k) \quad,
\label{eq:1a}
\ee
where $t$ is the dimensionless distance from criticality at zero
temperature ($T=0$), $k$
is the wavenumber, and $\Omega_n=2\pi Tn$ is a bosonic Matsubara frequency. 
$k$ and $\Omega_n$ have been made dimensionless by means of suitable
microscopic scales. The leading behavior of the coefficient $a$ for
small $k$ is
\be
a(k\rightarrow 0) \propto (\ln 1/k)^{-1/26}\quad,
\label{eq:1b}
\ee
\eml%
Such logarithmic corrections to power-law scaling can be conveniently expressed
in terms of scale dependent critical exponents. For instance, with $b\sim 1/k$
a RG length scale factor\cite{notation_footnote}, we can write 
$a(k)\,k^2 \propto k^{2-\eta}$, with a scale dependent critical exponent $\eta$ 
given by
\bml
\label{eqs:2}
\be
\eta = \frac{-1}{26}\,\ln\ln b/\ln b\quad.
\label{eq:2a}
\ee
The correlation length exponent $\nu$, the susceptibility exponent $\gamma$,
and the dynamical exponent $z$ can be directly read off Eqs.\ (\ref{eqs:1}).
The order parameter exponents $\beta$ and $\delta$ can be obtained from 
scaling arguments for the free energy. We find
\bea
\nu = 1/(2-\eta)\quad,\quad z=3-\eta\quad,
\nonumber\\
\gamma = 1\quad,\quad\beta = 1/2\quad,\quad\delta = 3\quad.
\label{eq:2b}
\eea
These exponents are defined as usual, i.e., $\xi\sim t^{-\nu}$,
$\Omega\sim T\sim\xi^{-z}$, ${\cal M}\sim t^{-\gamma}$, $m\sim t^{\beta}$, and
$m\sim h^{1/\delta}$, with $\xi$ the correlation length and $h$ an external 
magnetic field. The result for $\eta$ is valid
to leading logarithmic accuracy; the values of
$\gamma$, $\beta$, and $\delta$, as well as the relations between $\eta$ and
$\nu$ and $z$, respectively, are exact.
Finally, we define a specific heat exponent $\alpha$ by 
$C_{\rm V}\propto T^{-\alpha}$ at criticality. (This is a generalization of
the usual definition of $\alpha$ at thermal phase transitions.) We obtain
the exact relation
\be
\alpha = -1 - \eta/z - \ln\ln b/\ln b \quad.
\label{eq:2c}
\ee
\eml%

In $d=3-\epsilon$, the second order transition can be treated in an
$\epsilon$-expansion. We find
\be
\eta = -\epsilon/26\quad,\quad \alpha = -1 + (\epsilon - \eta)/z\quad.
\label{eq:3}
\ee
The value for $\eta$ is valid to one-loop order, the second equality
is exact. The other exponents are still given by the exact
Eqs.\ (\ref{eq:2b}) above.

These results predict that the transition in $3-d$, to the extent that it is of
second order, is characterized by mean-field exponents with 
logarithmic corrections. Within the accuracy of existing experiments, this is
indistinguishable from Hertz's critical behavior. In addition to explaining
why the transition is of first order in some materials, and of second order
in others, our theory therefore provides an explanation for the fact that
the observed critical behavior in the second order case is mean field-like.
We now sketch the derivation of the above results.

The lesson learned from the disordered quantum ferromagnetic 
problem\cite{us_fm_dirty} is the following:
For a reliable analysis of the critical behavior it does not suffice to
construct a LGW theory. Rather, in addition to the order parameter fluctuations,
all other soft modes that couple to the latter must be
kept explicitly and on equal footing. Accordingly, the effective action
should consist of a part depending on the order parameter field ${\bf M}$,
a part depending on the soft fermionic two-particle modes described by a 
field $q$, and a coupling between the two,
\be
{\cal A}[{\bf M},q] = {\cal A}_{M} + {\cal A}_q + {\cal A}_{M,q}\quad.
\label{eq:4}
\ee
${\cal A}_M$ is a static LGW functional (the dynamics will be provided by
${\cal A}_{M,q}$),
\be
{\cal A}_M = \int dx\ {\bf M}(x)\,[t - a\nabla^2]\,{\bf M}(x)
             + u\int dx\ {\bf M}^4(x).  
\label{eq:5}
\ee
Here $x\equiv (\bf{x},\tau)$ comprises the real space position ${\bf x}$ 
and the imaginary time $\tau$. $\int dx = \int d{\bf x}\int_0^{\beta} d\tau$ 
with $\beta = 1/k_{\rm B}T$. $a$ and $u$ are constants. 

The soft fermion field $q$ originates from the composite fermion 
variables\cite{us_fermions}
\bml
\label{eqs:6}
\be
Q_{12} = \frac{i}{2}\left(
\begin{array}{cccc}
-\psi _{1\uparrow }{\bar\psi }_{2\uparrow } & -\psi _{1\uparrow }
{\bar\psi }_{2\downarrow } & -\psi _{1\uparrow }\psi _{2\downarrow } & 
\psi _{1\uparrow }\psi _{2\uparrow } \\ 
-\psi _{1\downarrow }{\bar\psi }_{2\uparrow } & -\psi _{1\downarrow }
{\bar\psi }_{2\downarrow } & -\psi _{1\downarrow }\psi _{2\downarrow }
& \psi _{1\downarrow }\psi _{2\uparrow } \\ 
{\bar\psi }\hspace{0.02in}_{1\downarrow }{\bar\psi }_{2\uparrow }
& {\bar\psi }_{1\downarrow }{\bar\psi }_{2\downarrow } & {\bar
\psi }_{1\downarrow }\psi _{2\downarrow } & - {\bar\psi }_{1\downarrow
}\psi _{2\uparrow } \\ 
-{\bar\psi }_{1\uparrow }{\bar\psi }_{2\uparrow } & - {\bar
\psi }_{1\uparrow }{\bar\psi }_{2\downarrow } & - {\bar\psi }
_{1\uparrow }\psi _{2\downarrow } & {\bar\psi }_{1\uparrow }\psi
_{2\uparrow }
\end{array}\right)\ .
\label{eq:6a}
\ee
Here the $\psi$ and ${\bar\psi}$ are the Grassmann-valued 
fields that provide the basic description of the electrons,
and all fields are understood to be taken at position $\bf{x}$. The indices
$1$, $2$, etc. denote the dependence of the fields on fermionic
Matsubara frequencies $\omega_{n_1} = 2\pi T (n_1+1/2)$, etc., and the arrows
denote the spin projection. A convenient basis in the space of $4\times 4$
matrices is given by $\tau_r\otimes s_i$ ($r,i=0,1,2,3$), with 
$\tau_0 = s_0$ the
$2\times 2$ unit matrix, and $\tau_{1,2,3} = -s_{1,2,3} = -i\sigma_{1,2,3}$,
with the $\sigma_i$ the Pauli matrices.
The matrix elements of $Q$ are bilinear in the
fermion fields, so
$Q$-$Q$ correlation functions describe two-fermion excitations.
In a Fermi liquid, the $Q$ fluctuations are massive and soft,
respectively, depending on whether the two frequencies carried by
the $Q$ field have the same sign, or opposite signs.
We thus separate the $Q$ fluctuations into massless modes, 
$q$, and massive modes, $P$, by
splitting the matrix $Q$ into blocks in frequency space\cite{us_fermions}, 
\bea
Q_{nm}({\bf x}) = \Theta(nm)\,P_{nm}({\bf x}) &+& \Theta(n)\Theta(-m)\, 
                                                  q_{nm}({\bf x})
\nonumber\\
          &&\hskip -18pt +\Theta(-n)\Theta(m)\,q^{\dagger}_{nm}({\bf x})\quad.
\label{eq:6b}
\eea
\eml%
In what follows, we will incorporate the frequency constraints expressed
by the step functions into the fields $P$ and $q$, respectively. That is,
the frequency indices of $q$ must always have opposite signs.

The massive modes can be formally integrated out to obtain an effective
action for the soft modes, $q_{nm}$. The Gaussian part of the fermionic 
action has the form
\bml
\label{eqs:7}
\be
{\cal A}_{q} = \frac{-1}{G}\int\! d{\bf x}\,d{\bf y}\sum_{1,2,3,4}\tr\left(
         q_{12}({\bf x})\,\Gamma_{12,34}^{(2)}({\bf x}-{\bf y})\,
         q_{34}^{\dagger}({\bf y})\right),
\label{eq:7a}
\ee
where $\tr$ traces over the matrix degrees of freedom of Eq.\ (\ref{eq:6a}).
As we see from Eq.\ (\ref{eq:6a}), the $q$ propagator describes 
particle-hole excitations, which in a clean electron system have
a ballistic dispersion relation, i.e., the frequency scales linearly with
the wavenumber. The vertex function $\Gamma^{(2)}$ in momentum space
therefore has the form
\be
\Gamma_{12,34}^{(2)}(k) = \delta_{13}\,\delta_{24}\,
    (k + GH\vert\Omega_{1-2}\vert)\quad.
\label{eq:7b}
\ee
\eml%
$G$ and $H$ are model dependent coefficients. A 
spin-singlet interaction amplitude can be included in the model, but
will not be of qualitative importance for our purposes.

The coupling term ${\cal A}_{M,q}$ originates from the linear coupling
between ${\bf M}$ and the electron spin density, which can be expressed
in terms of $Q$ by means of Eq.\ (\ref{eq:6a}). It is obvious that
integrating out the massive field $P$ will result in terms that couple
${\bf M}$ with all powers of $q$, 
${\cal A}_{M,q} = {\cal A}_{M-q} + {\cal A}_{M-q^2} + \ldots$. 
Let us define a matrix magnetization field $B({\bf x})$ by
\bml
\label{eqs:7'}
\be
B_{12}({\bf x}) = \sum_{i,r} \left(\tau_r\otimes s_i\right)\,(-)^{r+1}\ 
                  {^i_rB}({\bf x}) \quad,
\label{eq:7'a}
\ee
with components
\be
{_r^i B}_{12}({\bf x}) = \sum_{n}\delta_{n,n_1-n_2}\left[
   M_n^i({\bf x}) 
 + (-)^{r+1} M_{-n}^i({\bf x})\right],
\label{eq:7'b}
\ee
\eml%
The first term in that series can then be written
\bml
\label{eqs:8}
\be
A_{M-q} = c_1T^{1/2}\int d{\bf x}\ \tr\left(B({\bf x})\,q({\bf x})\right)
           \quad.
\label{eq:8a}
\ee
The second one has the overall form
\be
{\cal A}_{M-q^2} \approx c_2 T^{1/2}\int d{\bf x}\ \tr 
   \left(B({\bf x})\,q({\bf x})\,q^{\dagger}({\bf x})\right)\quad.
\label{eq:8b}
\ee
\eml%
Its detailed structure will be derived 
elsewhere\cite{us_tbp}. $c_1$ and $c_2$ are model dependent
coupling constants.

The Gaussian field theory defined by the terms bilinear in ${\bf M}$ or
$B$ and $q$ is easily diagonalized in terms of the paramagnon propagator
\bml
\label{eqs:9}
\be
{\cal M}(k,\Omega_n) = 1/\left(t + ak^2 + \frac{(4G c_1^2/\pi)\vert
    \Omega_n\vert}{k + GH\vert\Omega_n\vert}\right)\quad,  
\label{eq:9a}
\ee
and the fermion propagator
\be
{\cal D}(k,\Omega_n) = 1/(k + GH\vert\Omega_n\vert)\quad,
\label{eq:9b}
\ee
\eml%

We now subject the entire action to a RG analysis\cite{Ma}.
We employ a differential momentum-shell RG and
integrate over all frequencies. With $b$ the RG length rescaling factor,
we rescale wavenumbers and the fields via
\bml
\label{eqs:10}
\bea
k &\rightarrow&bk\quad,
\label{eq:10a}\\
{\bf M}_n({\bf x})&\rightarrow& b^{(d-2+\eta)/2}{\bf M}_n({\bf x})\quad,
\label{eq:10b}\\
q_{nm}({\bf x})&\rightarrow& b^{(d-2+{\tilde\eta})/2}q_{nm}({\bf x})\quad.  
\label{eq:10c}
\eea
Here ${\bf M}_n({\bf x})$ is the temporal Fourier transform of ${\bf M}(x)$, 
and $\eta$ and ${\tilde\eta}$ are exponents that characterize the spatial
correlations of the order parameter and the fermion fields, respectively.
The rescaling of imaginary time, frequency, or temperature is less
straightforward. We need to acknowledge the fact that there are two
different time scales in the problem, namely, one that is associated
with the critical order parameter fluctuations, and one that is associated
with the soft fermionic fluctuations. Accordingly, we must allow for two
different dynamical exponents, $z$ and ${\tilde z}$, and imaginary time and
temperature may get rescaled via either one of two possibilities,
\bea
\tau \rightarrow b^{-z}\tau\quad,\quad T \rightarrow b^{z} T\quad,
\label{eq:10d}\\
\tau \rightarrow b^{-{\tilde z}}\tau\quad,\quad 
                                T \rightarrow b^{{\tilde z}} T\quad,
\label{eq:10e}
\eea
\eml%
How these various exponents should be chosen is discussed 
below.

Within this framework, Hertz's theory corresponds to a fixed point where
${\tilde\eta} = {\tilde z} = 1$, which makes $G$ and $H$ marginal, and
$\eta = 0$, $z = 3$, which makes $a$ and $c_1$ marginal. Power counting
then shows that $c_2$ is irrelevant for $d>1$ if the time scale is given
by the exponent $z$, but marginal for $d=3$ and relevant for $d<3$ if it
is given by ${\tilde z}$. It is easy to find explicit diagrams, starting
at one-loop order, where the latter is the case\cite{us_tbp}. 
This establishes that
the upper critical dimension, above which Hertz's fixed point is stable,
is $d_c^+=3$.

To deal with the situation in $d\leq 3$ we go to one-loop order.
Perturbation theory, combined with power counting, shows that in 
$d=3$ there are no logarithmic corrections to $c_1$, $c_2$, $G$, and $t$.
Motivated by the disordered case\cite{us_fm_dirty}, we will be looking
for a fixed point where $G$ and $c_1$ are marginal, which implies 
${\tilde\eta}=1$ and $z + \eta = 3$. As mentioned above, $c_2$ 
can have two different scale dimensions, depending on which time scale
enters. We require $c_2$ with the fermionic time scale, corresponding
to ${\tilde z}$, to be marginal, which yields ${\tilde z} + \eta = 1$.
This leaves $\eta$ as the only independent scale dimension.
The critical time scale makes $c_2$ irrelevant with scale dimension
$-1$. For the remaining coupling constants to one-loop order we find
the flow equations
\bml
\label{eqs:11}
\bea
da/d\ln b &=& -\eta\,a - A_a/H\quad,
\label{eq:11a}\\
du/d\ln b &=& -(2+\eta)u - A_u c_2^2/H\quad,
\label{eq:11b}\\
dH/d\ln b &=& \eta\,H + A_H/(a+t)\quad.
\label{eq:11c}
\eea
Here $c_2$ is the irrelevant version of $c_2$, with flow equation
\be
\hskip -82pt dc_2/d\ln b  =  -c_2\quad.
\label{eq:11d}
\ee
\eml%
$A_u>0$ is a model dependent positive coefficient.
An explicit calculation yields for the other two coefficients, 
$A_H = 27 A_a = 3Gc_2^2/\pi^3$, with $c_2$ the marginal incarnation
of this coupling constant. The ratio $A_H/A_a = 27$ is what determines
the critical exponents.

Solving the flow equations, Eqs.\ (\ref{eqs:11}), at $t=0$ shows that $u$ 
becomes negative at a finite scale provided that its bare value satisfies
$u^{(0)} < A_u\,(c_2^{(0)})^2/A_a$. This results in a fluctuation-driven 
first order phase transition. Indeed, the generalized mean-field theory
for this transition maps onto the one for the
superconducting transition at $T>0$, which is the canonical example of
a fluctuation-driven first order transition\cite{HalperinMa}. However,
if the opposite inequality holds, then $u$ remains positive at all scales
and the transition is continuous. Notice that within strict perturbation
theory the second term on the right-hand side of Eq.\ (\ref{eq:11b}) is
constant, so the transition is always of first order. This is the RG
version of the theory given in Ref.\ \onlinecite{us_letter}. We note that
for sufficiently large $t$, perturbation theory is valid. If the first
order transition predicted by perturbation theory occurs at
sufficiently large $t$, it therefore is unaffected by the fluctuation
effects. In order for the generalized mean-field theory of
Ref.\ \onlinecite{us_letter} to be controlled, the magnetization
discontinuity must in addition be small.

As we see, fluctuations can qualitatively change the prediction of perturbation 
theory and drive the transition second order. The mechanism for this is very
similar to the fluctuation-driven second order transition in classical Potts
models\cite{FucitoParisi}. The point
is that the renormalization of a negative loop correction to the $u$-flow 
equation can make this term go to zero for large scales sufficiently fast to
keep $u$ positive, even if a scale independent loop correction would lead to
a negative $u$. In the case of a second order transition, the critical 
exponents $\eta$, $\nu$, and $\gamma$
can be obtained by solving Eqs.\ (\ref{eqs:11}) and substituting the
result for $a(\ln b)$ in the paramagnon propagator, Eq.\ (\ref{eq:9a}).
In $d=3$ this procedure is tricky since it leads to scale dependent exponents;
technical details will be reported elseswhere\cite{us_tbp}. The results are
given in Eqs.\ (\ref{eqs:2}). In $d=3-\epsilon$ the procedure is 
straightforward and leads to $\eta$ as given in Eqs.\ (\ref{eq:3}), and
to Eqs.\ (\ref{eq:2b}).
$\alpha$ is most easily obtained from the Gaussian free energy
by replacing the Gaussian paramagnon propagator by the critical one.
Differentiation with respect to $T$ yields the specific heat. 
$\beta$ and $\delta$ are most easily obtained from scaling
arguments\cite{us_tbp}, taking into account that $u$ is a 
dangerous irrelevant operator for the magnetization\cite{Ma}.


This work was supported by the NSF under grant Nos. DMR-01-32555 and 
DMR-01-32726.

\vfill\eject
\end{document}